\documentstyle[epsf,aps,preprint]{revtex}

\newcommand{\cp}{C\!P}
\newcommand{\lsim}{\mathrel{\lower4pt\hbox{$\sim$}}
\hskip-12.5pt\raise1.6pt\hbox{$<$}\;}
\newcommand{\gsim}{\mathrel{\lower4pt\hbox{$\sim$}}
\hskip-12.5pt\raise1.6pt\hbox{$>$}\;}

\begin{document}
\baselineskip14pt

\vspace{-.5in}
\hskip4.5in
\vbox{\hbox{BNL-HET-99/40}
\hbox{OITS-683}
\hbox{hep-ph/9911419}}

\vspace{.5in}
\begin{center}
{\large\bf Novel  $CP$-violating Effects in $B$ decays from 
Charged-Higgs   \\ \medskip  in a
Two-Higgs Doublet Model for the
Top Quark }
\vspace{.5in}

Guo-Hong Wu$^a$\footnote{Email: wu@dirac.uoregon.edu, $^\dag$soni@bnl.gov}
and Amarjit Soni$^{b\dag}$
\vspace{.3in}

{\it $^a$ Institute of Theoretical Science, University of Oregon,
        Eugene, OR 97403-5203}

\medskip

{\it $^b$ Department of Physics,
Brookhaven National Laboratory, Upton, NY 11973-5000 }

\vspace{.3in}


\end{center}

\bigskip

\baselineskip24pt

\begin{abstract} 
\addtolength{\baselineskip}{-.3\baselineskip}
We explore charged-Higgs $\cp$-violating effects in a specific 
type III two-Higgs
doublet model which is theoretically attractive as it accommodates
the large mass of the top quark in a natural fashion.
Two new $CP$-violating phases arise from the right-handed
up quark sector.
We consider $CP$ violation in both neutral and charged $B$ decays.
Some of the important findings are as follows.
{\bf 1)} Large direct-$CP$ asymmetry is found to be possible
for $B^\pm \to \psi/J \  K^\pm$.
{\bf 2)} Sizable $D\bar{D}$ mixing effect at the percent level
is found to be admissible
despite the stringent constraints from the
data on $K\overline{K}$ mixing,
$b \to s \gamma$ and $B^- \to \tau \bar{\nu}$ decays.
{\bf 3)} A simple but distinctive $CP$ asymmetry pattern emerges in
decays of $B_d$  and $B_s$ mesons, including
$B_d \to  \psi/J \ K_S, \ D^+D^-$, and  
$B_s \to D^+_s D^-_s, \ \psi \  \eta/\eta^\prime, \ \psi/J \ K_S$. 
{\bf 4)} The effect of $D\bar{D}$ mixing on the $CP$ asymmetry 
in $B^\pm \to D/\bar{D} K^\pm$ and on the extraction of the
angle $\gamma$ of the unitarity triangle from 
such decays can be significant. 

\end{abstract}

\newpage


\section{Introduction}

 One of the main programs at the upcoming B factories is to measure
the size of CP violation in as many $B$ decay modes as possible 
so as to establish the pattern of CP violation among various B decays
\cite{BaBar}.
This then may allow for an experimental test not only of the standard model
(SM) Cabibbo-Kobayashi-Maskawa (CKM) paradigm for CP violation,
but also many extensions of the SM  that often contain new sources of
CP violation. 

  In this paper, we continue to study \cite{ksw}
 the distinctive phenomenological implications of 
a two Higgs doublet model for the top quark that
is designed to take into account the large mass of the top quark in a
natural fashion.
This model  contains flavor violation and 
new sources of $CP$ violation in the charged Higgs sector.
In this top-quark two-Higgs doublet model (T2HDM) first introduced in 
Ref.\cite{daskao}, the top quark is assigned a special status by 
coupling it to one Higgs 
doublet that gets a large vacuum expectation value (VEV), 
whereas all the other quarks are coupled
only to the other Higgs doublet whose VEV is much smaller.
  This arrangement of Yukawa interactions is motivated by the 
mass hierarchy between the top and the other quarks, and the T2HDM
can be considered as a special case of the general 2HDM 
(type III)\cite{type3}.
The unique predictions of the model for the $CP$ asymmetries in both
neutral and charged $B$ decays should allow for many experimental
tests at the $B$-factories.

 One notable feature about type III 2HDMs is the fact that 
natural flavor conservation (NFC) \cite{NFC} is not imposed on the
Yukawa interactions as is done in models I and II of 2HDM \cite{hhunter}.
However, the assumption of NFC is more of a convenience than a necessity
\cite{type3}, and this is especially true for the top quark as at the moment
there is no experimental data that require it.
Relaxing the assumption of NFC 
leads to many interesting  phenomenological implications 
\cite{type3}.
As a result,  three distinctive features arise in this 
T2HDM which are absent in models with NFC.
Firstly, there are new CP-violating phases (in the charged Higgs sector)
 besides the CKM phase.
 These new phases come from the unitary diagonalization matrix
acting on the right-handed (RH) up-type quarks.
Secondly, some charged Higgs Yukawa couplings are greatly enhanced
by the large ratio of the two Higgs VEVs 
denoted by $\tan \beta$.
This is the case, for example, in 
$H^+\bar{c}_Rq_L$ ($q=d,s,b$) and $H^+\bar{u}_Rb_L$, whereas these couplings
are suppressed by $1/\tan \beta$ in models with NFC.
These two features have important implications for
$K\bar{K}$ and $D\bar{D}$ mixing, as well as for CP violation in $B$ decays.
Finally, flavor changing neutral Higgs (FCNH) couplings exist
among the up-type quarks but not the down-type quarks,
and could contribute, for example, to $D\bar{D}$ mixing at tree level
\cite{daskao}. 

  It is the first two aspects of the model that we wish to concentrate on 
in this work. An immediate consequence of these two features is 
 the resulting complex tree-level $b\to c$ decay amplitudes.
This distinguishes it from many popular models with
new physics at the loop level, and it
has important implications both for neutral $B$ decays to $CP$ eigenstates
and for direct $CP$ violation in charged $B$ decays.
 In a previous note \cite{ksw}, we have highlighted the implications
of the model for the $CP$ asymmetry in the ``gold-plated" mode
$B\to \psi/J K_S$ and found that the asymmetry 
could take very different values from the SM expectation. 
In this work, we would like to extend our previous analysis in two ways. 

First, we will perform a systematic analysis of the $CP$ asymmetries
in the various $B_d$ and $B_s$ decay channels, including, for example, 
$B_d \to \psi/J \  K_S, \  D^+D^-, \  \pi\pi$, and  
$B_s \to D^+_s D^-_s, \  \psi \  \eta/\eta^\prime , \  \psi/J \  K_S$ .
A simple and distinctive pattern of $CP$ asymmetry emerges 
from our study. 
As a result, studies of the $CP$ asymmetries in a few $B$ decay modes 
may be utilized to extract information on both 
 the angle $\beta_{\rm CKM}$\footnote{To avoid confusion with the
$\beta=\tan^{-1} v_2/v_1$ associated with the ratio of Higgs VEVs in the 2HDM,
we use $\beta_{\rm CKM}$ to denote one of the angles of the 
the CKM unitarity triangle.} 
of the CKM  Unitarity Triangle (UT) 
and one new $CP$-violating phase of this model.
With more measurements, one can confirm
or rule out the model via consistency checks. 
 
Second, on account of the new $CP$ phase in the Higgs-mediated decay 
amplitudes, 
we investigate in the T2HDM direct $CP$-violating effects
in charged $B$ decays, including $B^\pm \to \psi/J K^\pm$ and
$B^\pm \to D/\bar{D} K^\pm$.
The former mode could exhibit a sizable partial rate asymmetry (PRA)
if the strong phase difference is not small. 
This mode is of special interest due to its experimental
cleanliness and high branching ratio.
The $B^\pm \to D/\bar{D} K^\pm$ mode is of interest for
the measurement of the angle $\gamma$ of the unitarity triangle\cite{ADS}
within the SM where the  $D\bar{D}$ mixing effect is negligible.
One interesting implication of the T2HDM is that the  $D\bar{D}$ mixing
can be significantly enhanced so that 
$x_D\equiv \Delta m_D/\Gamma_D = {\cal O}(10^{-2})$.
Improvements in the existing bound on $x_D$ may therefore be very 
worthwhile.
This possible enhancement arises because the
$D\bar{D}$ mass difference is quite sensitive to the poorly constrained
$u_R-t_R$ mixing of the right-handed sector, and a large mass splitting 
due to charged Higgs can occur. 
This large $D\bar{D}$ mixing can in turn strongly modify the direct 
$CP$ asymmetry
in $B^\pm \to D/\bar{D} K^\pm$. The implication for the extraction of $\gamma$
within the T2HDM will be discussed.

  The outline of the paper is as follows: the model is introduced
in section II. The most stringent experimental constraints from 
$K\bar{K}$ mixing, $b \to s \gamma$ and $B \to \tau \bar{\nu}$  
decays are presented in section III.
In section IV, we explore the pattern of $CP$ asymmetries in
neutral $B_d$ and $B_s$ decays.
Section V is devoted to the study of direct $CP$ violation in
charged $B$ meson decays in the T2HDM, first in $B^\pm \to \psi/J K^\pm$,
then in $B^\pm \to D/\bar{D} K^\pm$.
We conclude with some discussion in section VI.

\section{The model}
\label{sec:model}

  The fact that the top quark mass is of order the weak scale and is much 
larger than the other five quarks
is suggestive of a different origin for its mass than the other five.
Many attempts have been made along this direction, including the dynamical
top-condensation model\cite{topCond} and  the top-color model\cite{topColor}.
As an alternative to the dynamical models of the top mass and electroweak
symmetry breaking, a top-quark 2HDM (T2HDM) was proposed \cite{daskao} to
accommodate the top mass and the weak scale through a separate Higgs doublet 
than the one that is responsible for the masses of the other quarks
and all the charged leptons.
The T2HDM can be viewed as an effective low energy parameterization 
through the Yukawa interactions
of some high energy dynamics which generates both the top mass and the 
weak scale. The two scalar doublets could be composite, as in top-color
models, and the Yukawa interactions could be the  residual effect of some 
higher energy four-Fermi operators.
Indeed, some similarities can be noted \cite{heyuan} 
in the effective low energy flavor
physics between dynamical top models and certain Model III 2HDMs including
the T2HDM.

   The Yukawa interaction of the T2HDM can be simply written as follows,
\begin{eqnarray}
        {\cal L}_Y & = & -\overline{L}_L\phi_1 E\ell_R
                -\overline{Q}_L\phi_1 Fd_R
                -\overline{Q}_L\widetilde{\phi}_1 G {\bf 1^{(1)}} u_R
                -\overline{Q}_L\widetilde{\phi}_2 G {\bf 1^{(2)}} u_R
                + {\rm H.c.},
        \label{eq:yuk}
\end{eqnarray}
where the two Higgs doublets are denoted by $\phi_i$ with 
$\widetilde{\phi}_i = i\sigma^2\phi_i^\ast$ $(i=1,2)$,
and where the  $3\times3$ Yukawa matrices  $E$, $F$ and $G$ give masses
respectively to the charged leptons, the down and up type quarks;
${\bf 1^{(1)}}\equiv {\rm diag}(1,1,0)$ and 
${\bf 1^{(2)}}\equiv {\rm diag}(0,0,1)$ are two orthogonal projection operators
onto the first two and the third families respectively,
and $Q_L$ and $L_L$ are the usual left-handed quark and lepton doublets.
The heaviness of the top quark arises as a result of 
the much larger VEV of $\phi_2$ to which no other quark couples. 
As a result,
one notable feature about this model is that the ratio of the two Higgs VEVs,
$\tan \beta=v_2/v_1$ ,
is required to be large so that the 
Yukawa couplings of the top and bottom quarks are
similar in magnitude. We will take $\tan\beta\ge 10$
 in the following analysis.

  This assignment of Yukawa couplings is not in line with the notion of NFC,
and this leads to two tree level FCNH interactions, 
$h\bar{t}c$ and $h\bar{t}u$, both of which are not constrained by present data.
Some phenomenological studies of the neutral Higgs sector, including 
the effect on $\bar{D}D$ mixing, can be found in~\cite{daskao}. 
On the other hand, only little attention has been paid to the charged Higgs 
sector\cite{ksw}, and it is this aspect of the
model on which we would like to focus.

 The charged Higgs Yukawa couplings can be obtained as,
\begin{eqnarray}
      {\cal L}^C_Y & = & \frac{g}{\sqrt{2}m_W}\left\{
      -\overline{u}_L V M_D d_R\left[G^+ -\tan\beta H^+\right]
      +\overline{u}_R M_U V d_L\left[G^+ -\tan\beta H^+\right] 
      \right. \nonumber \\
                & & \;\;\;\;\;\;\;\;\;\;\;\;\;\; \left.         
      +\overline{u}_R \Sigma^\dagger V d_L\left[\tan\beta +
                        \cot\beta\right]H^+ +{\rm H.c.}\right\},
        \label{eq:chiggs}
\end{eqnarray}
where $G^\pm$ and $H^\pm$ represent the would-be Goldstone bosons
and the physical charged Higgs bosons, respectively, and
 $M_U$ and $M_D$ are the diagonal up- and down-type mass matrices.
 Here $\Sigma \equiv M_U U_R^\dagger {\bf 1^{(2)}} U_R$, where
 $U_R$ denotes the unitary rotation of the RH up-type quarks
from gauge to mass eigenstates.


  One feature that distinguishes the T2HDM from 2HDM's with NFC 
(models I and II)  
is the presence of the unitarity matrix $U_R$ contained in the 
$\Sigma$ matrix. This gives rise to two new $CP$-violating phases
as well as non-standard Yukawa couplings.  
The unitary matrix $U_R$ can, in general,  be parameterized as:
\equation
        U_R = \left(\begin{array}{ccc}
         \cos\phi & -\sin\phi & 0 \\
         \sin\phi & \cos\phi & 0 \\
         0 & 0 & 1
                \end{array}\right)
              \left(\begin{array}{ccc}
         1 & 0 & 0 \\
         0 & \sqrt{1-|\epsilon_{ct}\xi|^2} & -\epsilon_{ct}\xi^\ast \\
         0 & \epsilon_{ct}\xi & \sqrt{1-|\epsilon_{ct}\xi|^2}
                \end{array}\right)
         \left(\begin{array}{ccc}
         \sqrt{1-|\epsilon_{ct}\xi^{\prime}|^2} & 0 &  
          -\epsilon_{ct}{\xi^{\prime}}^\ast \\
         0 & 1 & 0 \\
         \epsilon_{ct}\xi^\prime & 0 & \sqrt{1-|\epsilon_{ct}\xi^{\prime}|^2}
                  \end{array}\right),
\endequation
where $\epsilon_{ct}\equiv m_c/m_t$ and where $\xi$ and $\xi^\prime$ are
 complex numbers with $|\xi^\prime| < |\xi| = {\cal O}(1)$.
Note that the form of $\Sigma$ is independent of 
the 1-2 rotation and depends only on the two unknown complex parameters
$\xi=|\xi|e^{-i\delta}$ and $\xi^\prime=|\xi^\prime| e^{-i\delta^\prime}$.

\vskip 0.1 in

Table~1: Comparison of the magnitudes of 
charged Higgs Yukawa couplings in the T2HDM and 
   Models I and II 2HDM for large $\tan \beta$.
\vskip 0.1in
\begin{center}
\begin{tabular}{|c|ccc|}
\hline \hline
 ~~~~~~~~ vertices ~~~~~~~~
  & ~~~~~~~~~~~~~ T2HDM ~~~~~~~~~  &  ~~~~~~~~~~
  & ~~~Models I and II 2HDM ~~ \\ \hline\hline 
 $\overline{c}_R b_LH^+$  & $\simeq m_c \xi^* V_{tb} \tan\beta$ 
& $\gg$ & $m_cV_{cb} \cot \beta$ \\ \hline
  $\bar{c}_Rq_LH^+$ ($q=d,s$) &  
    $\simeq m_c \tan \beta (-V_{cq} + V_{tq}\xi^*)$
    & $\gg$ & $m_cV_{cq} \cot \beta$ \\ \hline
 $\bar{u}_Rb_LH^+$ & $\simeq m_c {\xi^\prime}^* V_{tb} \tan\beta$
 & $\gg$ & $m_uV_{ub}\cot{\beta}$ \\ \hline
 $\bar{t}_Rq_LH^+$ ($q=d,s,b$) & $\sim m_t V_{tq} \cot{ \beta}$ 
& $\sim$ & $m_tV_{tq} \cot {\beta}$    \\ \hline \hline
\end{tabular}
\end{center}

\vskip 0.1 in

  The phenomenologically important Yukawa couplings are listed in Table~1.
  Several remarks can now be made concerning Table~1.
In comparison to the popular 2HDM (model II) which is realized, e.g. in the
supersymmetric extension of the SM, the T2HDM contains
rather large $\overline{u}_R b_L H^+$,  $\overline{c}_R b_L H^+$,
$\overline{c}_R s_L H^+$, and  
$\overline{c}_R d_L H^+$ couplings, all enhanced by the large 
$\tan{\beta}$.
In particular, the anomalously large $\overline{u}_R b_L H^+$ coupling 
is directly proportional to $m_t V_{tb} \tan{ \beta}$ and to 
the $u_R-t_R$ mixing parameter $\epsilon_{ct} {\xi^\prime}^*$,
whereas in model II it depends on 
$m_u V_{ub} \cot {\beta}$ which is completely negligible. 
The unknown mixing parameter $\xi^\prime$ can be  constrained from 
the experimental data on $B^- \to \tau \bar{\nu}$ decay and 
$D\bar{D}$ mixing.
Similarly, the potentially large charm quark Yukawa interactions
can be expected to affect in a significant way
the $K\bar{K}$ system and the $CP$ asymmetries in $B$ decays.
By contrast, the charged Higgs contribution to $B\bar{B}$ mixing is negligible
due to the $1/\tan{\beta}$ suppression of the top quark 
Yukawa coupling.

 We now turn to a detailed analysis of the most stringent experimental 
constraints from $K\bar{K}$, $b\to s \gamma$, and $B^- \to \tau \bar{\nu}$. 
For the numerical estimates, we will assume $|\xi| = 1$ unless otherwise 
stated.

\section{Experimental Constraints}
\label{sec:constraints}

\subsection{Constraints from $K\bar{K}$ mixing}

  The $\Delta S=2$ effective Hamiltonian receives contributions from
box diagrams with virtual $W$ and $H$ bosons and up-type quarks.
The short distance contribution to the $K_L$-$K_S$ mass difference 
$\Delta m_K$ in the standard model mainly comes from the $c$ quark.
In the 2HDM that we are considering, the new contribution is dominated by the
$HHcc$ box diagram for $\tan {\beta} > 10$.
The leading term in the $\Delta S=2$ effective Hamiltonian from charged Higgs
exchange can be expressed as 
\equation \label{eq:HK}
H^{\Delta S=2}_{H^+} \simeq
\frac{G_F^2}{16\pi^2} \lambda_c^2 \eta_1^{\prime}
\frac{m_c^4 \tan^4{ \beta}}{m_H^2} (\bar{d}_L \gamma_{\mu} s_L)^2
 + {\rm H.c.} \; .
\endequation
where $\eta_1^{\prime}$ is the short-distance QCD correction factor, and
$\lambda_c=V_{cs}V_{cd}^* - \xi V^*_{td} V_{cs}- \xi^* V_{ts} V^*_{cd} 
 + |\xi|^2 V_{ts} V^*_{td}$. The last three terms in 
$\lambda_c$ are suppressed by $\lambda^2$ ($\lambda=|V_{us}|=0.22$)
 relative to the first
term and   can be neglected in the mass difference $\Delta m_K$.
By way of contrast, these terms are essential for the $CP$-violation parameter
$\epsilon_K$.

The charged Higgs contribution to $\Delta m_K$ can be easily obtained from
Eq.~(\ref{eq:HK}) by setting $\lambda_c  \rightarrow V_{cs}V_{cd}^*$.
The total short distance contribution to $\Delta m_K$ is given by~\cite{ksw},
\begin{eqnarray}
 \left( \Delta m_K \right)_{\rm SD}
 & = & \frac{G_F^2}{6\pi^2} f_K^2 B_K m_K \lambda_c^2
\times ( m_c^2 \eta_1 + \frac{m_c^4 \tan^4 {\beta}}{4 m_H^2}
 \eta^{\prime}_1 )
\end{eqnarray}
where the first term is from the SM and the second from Higgs exchange,
$f_K=160 \; \rm{MeV}$ is the kaon decay constant,
$B_K=0.87 \pm 0.14$ \cite{paganini,lattice} is the bag factor, 
and $\eta_1=1.38 \pm 0.53$ \cite{Buras} and
$\eta^{\prime}_1$ are the QCD corrections to the two box diagrams.
The SM top quark contribution is a few percent of the charm quark 
contribution and is not included in the above equation. Similarly
the contribution from $WHcc$ and other box diagrams is negligible 
in the large $\tan {\beta}$ limit and is not considered.

 Because of the large  uncertainties in $B_K$, $\eta_1$, $m_c$ and in
the long distance contribution, 
we have used the method described in \cite{paganini} to derive
the constraints on the model. The QCD correction factor $\eta_1^{\prime}$
is unknown, and we simply assign it the value of $\eta_1$ and allow them
to vary independently within their $1\sigma$ ranges in our error analysis. 
Assuming the long distance effect to be $30\%$ \cite{herrlich}
 of $\Delta m_K$, we get
\begin{eqnarray} \label{eq:mkbound}
m_H/\tan^2 {\beta} & > & 0.48 \;\rm{GeV}
\end{eqnarray}
 for $\tan {\beta} > 10$ and at the  $95\%$ C.L..  Note that the
$\tan^4{ \beta}$ enhancement in $\Delta m_K$ leads to a severe
lower bound on the Higgs mass for large $\tan {\beta}$. This is
 a unique feature of the T2HDM.

  Similar to $\Delta m_K$, the $CP$-violating parameter $\epsilon_K$  also 
receives a significant contribution from charged Higgs exchange,
which is again dominated by the $HHcc$ box diagram due to its
$\tan^4 {\beta}$ dependence.
However, the physics of $\epsilon_K$ differs from that of $\Delta m_K$
in three important aspects.
First, unlike $\Delta m_K$, $\epsilon_K$ is short-distance dominated 
and is theoretically under better control.
Second, as the leading, first term of $\lambda_c$ is real,
the sub-leading, 2nd and 3rd terms now become important for
CP violation.
In fact, the charged Higgs contribution to $\epsilon_K$ is directly
proportional to the $c_R-t_R$ mixing parameter $\xi$, whereas 
$\Delta m_K$ is independent of $\xi$ to a good approximation.
Third, within the SM,  $\epsilon_K$ 
depends to a large extent on the top quark box diagram, and 
also on the $ct$ and $cc$ box diagrams to a lesser extent.
In contrast, the short distance contribution to $\Delta m_K$  is 
predominantly due to the charm quark.

 The SM expression for $\epsilon_K$ is well understood and can be 
found, for example, in \cite{Buras}.
The dominant Higgs contribution can be obtained from Eq.~(\ref{eq:HK}), 
\begin{eqnarray}
\epsilon_K^H & = &  e^{i \frac{\pi}{4}} C_{\epsilon} B_K A \lambda^4
 \eta_1^{\prime} \sqrt{\rho^2 + \eta^2} \sin(\gamma + \delta) |\xi|
  \frac{(m_c \tan {\beta})^4}{4m_W^2 m^2_H}  
\end{eqnarray}
where $A=0.82 \pm 0.04$, $\rho$, and $\eta$ are the CKM parameters 
in the Wolfenstein parameterization\cite{wolfenstein},
$\gamma \equiv \tan^{-1} \eta/\rho$ is one of the angles of the 
unitarity triangle, and 
$C_{\epsilon}=\frac{G^2_F f^2_K m^2_W m_K}{6 \sqrt{2} \pi^2 \Delta m_K}
=3.78 \times 10^4$.

  As $\gamma$ is  basically a free parameter in this model, 
we can obtain bounds on the parameter
$Y \equiv  \sin (\gamma+\delta) |\xi|
\left( \frac{\tan{ \beta}}{20} \right)^4
 \left( \frac{200 \; \rm{GeV}}{m_H} \right)^2$ 
for any given value of  $\gamma$
by allowing $\sqrt{\rho^2 + \eta^2}$ to vary within its $1\sigma$ 
uncertainties derived from $b\rightarrow u e \nu$, which receives 
negligible contribution from Higgs exchange.
 It is interesting to note that charged Higgs exchange can be solely
responsible for $\epsilon_K$ if the CKM matrix is real (i.e. $\gamma=0^o$).
Using the method of \cite{paganini} for error analysis, we obtain 
at the $95\%$ C.L.  $0.08 <  Y < 0.39$ for the case of a real CKM matrix.
  If we assume that $\gamma$ takes its SM central value of $68^o$ 
\cite{paganini}, the constraint becomes
 $ -0.085 < Y  < 0.08 $.
 And for $\gamma=-45^o$, we get the bound
$0.14 < Y  < 0.65$\footnote{As will be discussed later, the CDF
result on $a_{\psi K_S}$\cite{sin2beta} tends to disfavor a large, 
negative~$\gamma$.}.
Note that unlike $\Delta m_K$,  the $\epsilon_K$ constraint
in the $m_H - \tan {\beta}$ plane depends on $|\xi|$ and $\delta$,
and it could be more stringent than the $\Delta m_K$ bound of
Eq.~(\ref{eq:mkbound}) \cite{ksw}.

\subsection{The $b \to s \gamma$ decay rate}

  We now turn to the constraints from $b$ decays.
  The inclusive radiative decay $b\to s \gamma$ 
has been studied in detail in \cite{ksw}. 
The main result of that analysis can be summarized as follows.

\begin{figure}[ht]      
\centerline{\epsfxsize 4 truein \epsfbox{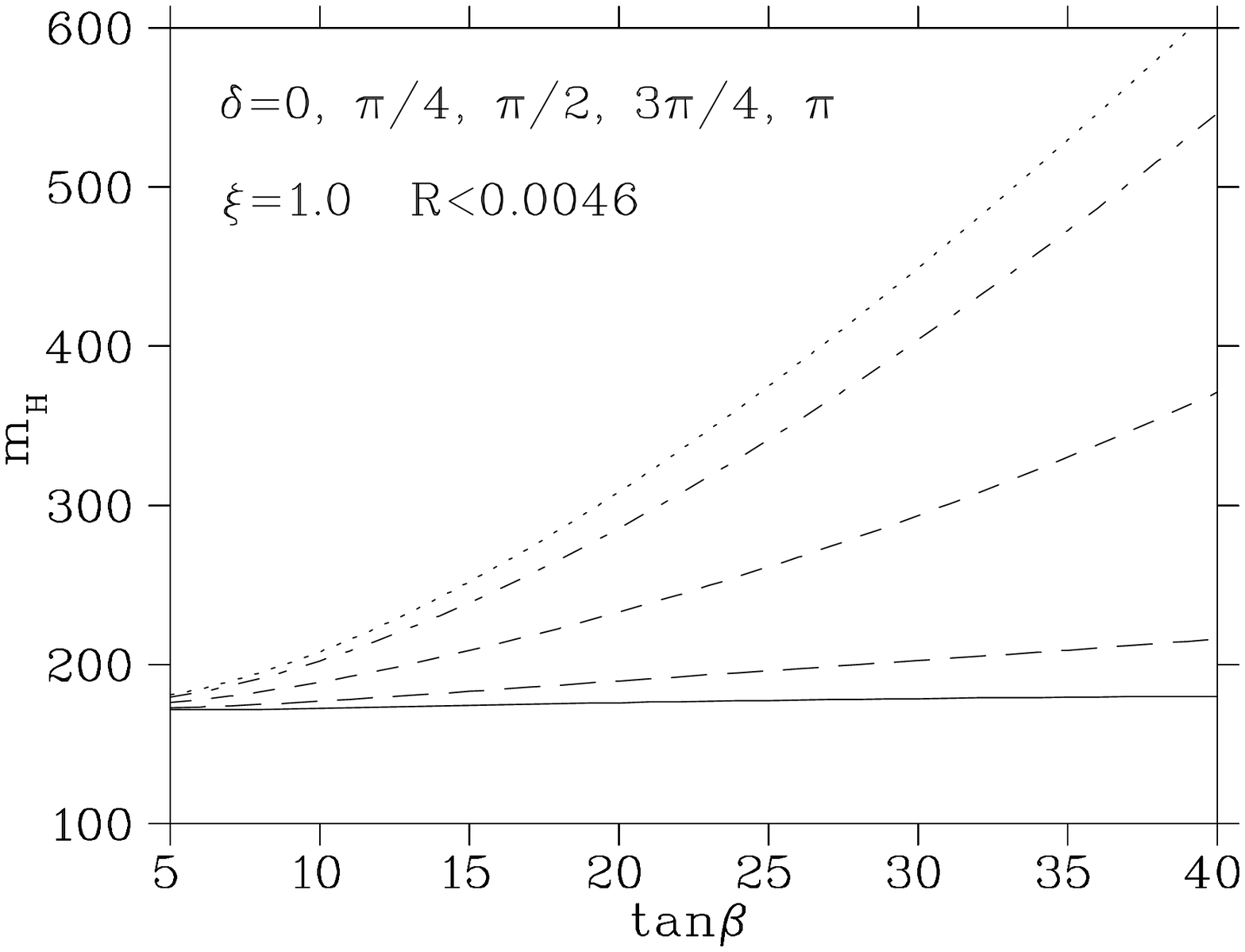}} 
\vskip .2 cm 
\caption[]{
\label{figbsg}
\small  
 The constraint from the $b\to s \gamma$ rate on the
 $m_H - \tan{\beta}$ plane of the T2HDM for $|\xi|=1.0$ 
and with different values of the phase $\delta$. 
The value of $\delta$ increases from the bottom curve to the top curve. 
The area below each curve is excluded.}
\end{figure}

As in Model II 2HDM,  a strong constraint is imposed from 
$b \to s \gamma$ on the allowed parameter space of the T2HDM.
However, the form of the constraint differs significantly.
 In model II, a lower bound of about $370$ GeV can be placed
on the charged Higgs mass, independent of $\tan{\beta}$
\cite{ciuchini}. 
A similar situation does not occur for the T2HDM
due to the nonstandard Higgs Yukawa couplings and the presence of 
the new $CP$-violating phase $\delta$.
In the T2HDM,  the charged Higgs amplitude
could interfere either constructively or destructively with the SM
amplitude depending on the new phase $\delta$.
Consequently, the lower limit on the
charged Higgs mass shows a strong dependence on both 
$\tan {\beta}$ and the phase $\delta$, and a much lower Higgs mass 
is still allowed.

Neglecting the charged-Higgs induced scalar operator
$(\bar{c}_Rb_L)(\bar{s}_Lc_R)$,
 we can derive  a bound on the             
ratio $R = {\cal B}(b\to X_s \gamma)/
                {\cal B}(b\to X_c e\overline{\nu})$
by combining the theoretical and experimental uncertainties
\cite{ksw}:
\equation
        0.0012 \leq R^{\rm theory} \leq 0.0046 ,
\endequation
where $R^{\rm theory}$ denotes the SM plus charged Higgs contribution.
The constraint on the $m_H-\tan {\beta}$ plane depends on both
$|\xi|$ and its phase $\delta$, and is shown in Fig.~\ref{figbsg}.

\subsection{$B^- \to \tau \bar{\nu}$}

 As noted in section~\ref{sec:model},
  the decay $B^- \to \tau \bar{\nu}$ could receive its dominant
contribution from charge Higgs exchange on account of the 
anomalously large $H^+ \bar{u}_R b_L$ coupling. 
Although the current experimental limit on 
$BR(B^- \to \tau \bar{\nu})$ is about one order of magnitude above the 
SM prediction,  it already starts to constrain the T2HDM in a
nontrivial way.

 The leading term of the effective Hamiltonian from charged
Higgs exchange that mediates this decay is given by
\begin{eqnarray}
{\cal H}_{eff}^H & = & 2\sqrt{2} G_F V_{tb}
  \frac{m_{\tau} m_c \tan^2{ \beta} {\xi^\prime}^*}{m^2_H}
  (\bar{u}_R b_L) (\bar{\tau}_R \nu_L)
 + \rm{H.c.}
\end{eqnarray}
  Assuming that the charged-Higgs-exchange contribution dominates over the SM
$W$-exchange, one can obtain the branching ratio as,
\begin{eqnarray}
BR(B^- \to \tau \bar{\nu}) & = & 
\tau_B \frac{G_F^2}{8\pi} f_B^2 m_{\tau}^2 m_B 
 \left( 1 - \frac{m_{\tau}^2}{m_B^2} \right)^2
 \left( \frac{m_B^2 m_c |\xi^\prime| \tan^2{ \beta}}{m_H^2 (m_b + m_u)}
 \right)^2   \nonumber \\
 & = & 4.6 \times 10^{-2} \left(\frac{f_B}{200\; \rm{MeV}}\right)^2 
  \left( \frac{10\; {\rm GeV}}{m_H/\tan{ \beta}} \right)^4 
   |\xi^\prime|^2
\end{eqnarray}
where $\tau_B$ is the $B$ lifetime, and 
we have used $f_B\simeq 200\;{\rm MeV}$ \cite{lattice}
 for the $B^-$ decay constant,
$m_c=1.3\;{\rm GeV}$, and $m_b=4.5\;{\rm GeV}$.
  The present experimental limit \cite{pdg}
$BR(B^- \to \tau \bar{\nu}) < 5.7 \times 10^{-4}$ then translates
into an upper bound on the $u_R - t_R$  mixing parameter $\xi^\prime$,
\equation \label{eq:Btaunubound}
 |\xi^\prime| < \left( \frac{m_H/{\rm GeV}}{30 \tan{ \beta}} \right)^2 \, .
\endequation

\subsection{$D\bar{D}$ mixing} 
\label{sect:DD}

 The SM effect on $D\bar{D}$ mixing is vanishingly small, 
estimated to be
$\Delta m_D \sim  {\cal O}(10^{-16}) \; {\rm GeV}$ \cite{golo}.
This is three orders of magnitude below the current experimental limit
 of $\Delta m_D < 1.6 \times 10^{-13} \; {\rm GeV}$ \cite{pdg}, 
which translates into an upper bound on the mixing parameter
$x_D\equiv \Delta m_D/\Gamma_D <0.1$.
As will be shown below, one loop
 charged Higgs exchange in the T2HDM could contribute to $x_D$
at the few percent level.

We note first that there exists in the T2HDM $\Delta C=2$ 
four-quark operators from tree level neutral Higgs exchange. 
This effect, however, is greatly suppressed by the RH 
mixings $|\epsilon^2_{ct} \xi\xi\prime|^2 \propto m^4_c/m^4_t$\cite{daskao},
\equation
 \left. \Delta m_D \right|_{h^0} \propto G_F (\tan{ \beta})^2 
m_c^6/m_t^4 m_h^2 \; .
\endequation
  In comparison, the one-loop box diagram with internal charged Higgs
and $b$ quarks is enhanced by the anomalously large Yukawa couplings
$H^+\bar{c}_Rb_L$ and $H^+\bar{u}_Rb_L$, both of which 
suffer no CKM suppression and grow as $\tan{\beta}$.
This feature differs significantly from the SM box diagram 
where the virtual $b$ quark effect, being highly Cabibbo-suppressed,
 is negligible compared to that of the $s$ and $d$ quarks.
 Simple power counting then gives for the $HHbb$ box diagram,
\equation
 \left. \Delta m_D \right|_{H^+} \propto G^2_F (\tan {\beta})^4 
m_c^4/ m_H^2 \; .
\endequation
This effect could be three orders of magnitude above
 the tree-level neutral Higgs contribution for the same choice of 
the mixing parameters, and can generate a sizable $x_D$ near the
threshold of experimental discovery. 

The calculation of the one-loop box diagram can be greatly simplified
by setting the external momenta of the $c$ and $u$ quarks to zero.
Neglecting the short-distance QCD correction,
the effective Hamiltonian can be obtained as,
\equation
 H^{\Delta C=2}_{H^+} \simeq 
\frac{G_F^2}{16\pi^2} (\xi {\xi^{\prime}}^*)^2 
\frac{m_c^4 \tan^4{ \beta}}{m_H^2} (\bar{u}_R \gamma_{\mu} c_R)^2 
+ {\rm H.c.} \; ,
\endequation
where we have taken $V_{tb}=1$.
  By use of the vacuum saturation approximation, one has
\begin{eqnarray}
 x_D & \simeq  & \frac{G_F^2}{6\pi^2} |\xi {\xi^{\prime}}^*|^2
 f_D^2 m_D  \frac{m_c^4 \tan^4 {\beta}}{4 m_H^2 \Gamma_D}  \\
 & \simeq & 7.5 \% \times
      |\xi {\xi^{\prime}}^*|^2 
    \left(\frac{1 \; {\rm GeV}}{m_H/\tan^2 {\beta}}\right)^2 \nonumber \\
 & < & 7.5 \% \times |\xi|^2 \left(\frac{m_H}{900 \; {\rm GeV}}\right)^2
\nonumber
\end{eqnarray}
where we have taken $f_D \simeq 0.2  \; {\rm GeV}$ \cite{lattice},
 and the $B\to \tau \nu$ bound 
(Eq.~(\ref{eq:Btaunubound})) has been imposed in the last step.
  Numerically,  one has $x_D \le 2\% |\xi|^2$ for 
$m_H=450\;\rm{GeV}$ and $\tan{\beta}=30$,
and $x_D \le 6\% |\xi|^2$ for
$m_H=800\;\rm{GeV}$ and $\tan\beta=40$. Both are consistent with the
$\Delta m_K$ and $b \to s \gamma$ constraints.  The $\epsilon_K$ constraint
can also be satisfied by choosing the phases of $\gamma$ and $\delta$,
whereas $x_D$ is independent of the phases.
Thus, in this model $x_D$ can be of order a few percent.
Continual experimental improvements over the existing bound
of $0.1$\cite{pdg} are therefore strongly encouraged.

   The neutral-meson mixing parameter $p/q$\cite{pdg} 
is also modified in the T2HDM.
To a good approximation, it is given by
\begin{eqnarray}
\frac{p_D}{q_D} & = & e^{-i2\theta_D} \; ,
\end{eqnarray}
where $\theta_D={\rm arg}(\xi) -{\rm arg}(\xi^\prime)$ is the mixing phase.
Recall that the mixing phase is zero in the SM.
 Finally, we note that the parameter $y_D=\Delta \Gamma/\Gamma_D$ is
very small in the T2HDM, as the $D/\bar{D}$ decay amplitudes
are unaffected by the Higgs exchange. 

\section{$CP$ asymmetry in neutral $B$ decays to $CP$ eigenstates}

\subsection{General Framework}

The non-standard Higgs interactions which we have 
been discussing could have significant effects on the 
time-dependent $\cp$-asymmetry
\equation \label{eq:asymdef}
       a(t) = \frac{\Gamma(B^0(t)\to f )-
                       \Gamma(\overline{B}^0(t)\to \bar{f})}
                       {\Gamma(B^0(t)\to f)+
                       \Gamma(\overline{B}^0(t)\to \bar{f})},
\endequation
where $\Gamma(B^0(t)\to f)$ ($\Gamma(\overline{B}^0(t)\to \bar{f})$)
 represents the time-dependent probability
for a state tagged as a $B^0$ ($\overline{B}^0$) at time $t=0$ to 
 decay into the final state $f$ ($\bar{f}$) at time $t$.
We will first consider the case when the final state is a
$CP$ eigenstate, and thus $\bar{f}=f$.

 It is convenient to introduce the re-phasing invariant quantity~\cite{pdg},
\equation
 \tilde{\lambda} \equiv \frac{q}{p} \frac{\overline{\cal A}}{{\cal A}}
\endequation
where ${\cal A} \equiv <f\ |B>$ and $\overline{\cal A} \equiv <f\ |\bar{B}>$
denote the $B$ and $\overline{B}$ decay amplitudes.
Neglecting the small $CP$-violating effect in $B\bar{B}$ mixing,
the neutral $B$ meson mixing parameter can be written as a pure phase:
$q/p=e^{-i2\phi_M}$ with $\phi_M$ the mixing phase.
The time-dependent $CP$ asymmetry can now be expressed in terms of
$\tilde{\lambda}$,
\equation  \label{eq:atlamb}
       a(t) = \frac{
 (1-|\tilde{\lambda}|^2) \cos (\Delta M t)
 -2 {\rm Im}(\tilde{\lambda}) \sin (\Delta M t) }
 {1 + |\tilde{\lambda}|^2} \, .
\endequation

   For certain channels without direct $CP$ violation,
i.e. $|\overline{\cal A}/{\cal A}|=1$, we can write
 $\overline{\cal A}/{\cal A}=f_{CP} e^{-2i\phi_D}$, with $\phi_D$ the 
phase of the decay amplitude and $f_{CP}=\pm 1$ the $CP$ eigenvalue 
of the final state $f$.
The $CP$ asymmetry then takes the 
simple form 
\equation \label{eq:atsimple}
       a(t) = - {\rm Im}(\tilde{\lambda}) \sin (\Delta M t)
            =  f_{CP} \sin 2(\phi_D + \phi_M) \sin (\Delta M t) \; .
\endequation

As we have noted, $B\bar{B}$ mixing receives a very small contribution
from charged Higgs exchange for moderate values
of $\xi$ after imposing the $K\bar{K}$ constraints.
We therefore only need to consider the charged Higgs effect
on the decay amplitudes.
The tree-level Hamiltonian for the transition 
$b\to c\bar{c}q$ ($q=s,d$) has a simple form in the large
$\tan {\beta}$ limit~\cite{ksw},
\equation
        {\cal H}_{\rm eff} \simeq 2\sqrt{2} G_F V_{cb}V_{cq}^\ast
                \left[ \overline{c}_L\gamma_\mu b_L
                        \overline{q}_L\gamma^\mu c_L
                +2\zeta e^{i\delta}
                \overline{c}_Rb_L\overline{q}_Lc_R\right] +{\rm H.c.},
        \label{eqn:heff}
\endequation
where
\equation
        \zeta e^{i\delta} \equiv \frac{1}{2}\frac{V_{tb}}{V_{cb}}
                \left(\frac{m_c\tan{\beta}}{m_H}\right)^2\xi^\ast,
                \label{defzeta}
\endequation
with $\zeta$ taken to be real and positive. 
The charged Higgs contribution, though suppressed by small Yukawa couplings,
 is enhanced by the CKM factor $V_{tb}/V_{cb} \simeq 25$ relative to 
the $W$-exchange amplitude.  As a result, $\zeta$ as large as $0.2$ is
allowed  by data~\cite{ksw}.
This could  significantly modify the SM predictions for the $CP$ asymmetries
in a variety of $B$ decay channels.

 The evaluation of  the Higgs amplitude  can be 
greatly simplified by assuming factorization,
which in this instance may be qualitatively reasonable due to the 
presence of a heavy quark in the initial and in the final state.
 The total amplitude can then be written as
\equation \label{eq:totamp}
{\cal A} \equiv {\cal A}({B}^0\to f) \simeq {\cal A}_{\rm SM}
 \left[ 1 - \zeta e^{-i\delta}\right.] \, .
\endequation
The ratio of the $B$ and $\bar{B}$ decay amplitudes is 
\equation \label{eq:ampratio}
\frac{\overline{\cal A}}{\cal A} = \frac{\overline{\cal A}_{\rm SM}}
        {{\cal A}_{\rm SM}}\exp(- 2 i \theta),
\endequation
with the new phase angle given by
\equation 
  \tan\theta = \frac{\zeta\sin\delta}{1-\zeta\cos\delta} .
\endequation
The allowed range for $\theta$ could be of order ten degrees for 
$\xi$ of order one \cite{ksw}.

  Therefore, in the T2HDM, the $CP$ asymmetries for neutral $B$ decays
to $CP$ eigenstates can be expressed in the general form,
\begin{eqnarray} \label{eq:atxi}
 a(t)  &   =  &
 f_{CP}
\sin 2(\phi_M^{\rm SM} + \phi_D^{\rm SM} + \theta) \sin (\Delta M t) \, ,
\end{eqnarray}
where direct $CP$-violating effect has been neglected.
The task for  $CP$ asymmetry study in the T2HDM is thus 
reduced to the evaluation of $\theta$.
Using Eq.~(\ref{eq:atxi}), we now examine the $CP$ asymmetry
pattern in $B_d$ and $B_s$ decays.

\subsection{$b\to c\bar{c}s$}

 In the SM, the quark transition $b\to c\bar{c}s$ proceeds
mainly through tree-level $W$ exchange. 
Even including the penguin contribution, the amplitude for
 $b\to c\bar{c}s$ 
is proportional to a single CKM phase ${\rm arg}(V_{cb}V^*_{cs})$ to a very
good approximation.
Direct $CP$ violation is therefore absent in this process and
$\left|\overline{\cal A}\right| 
=\left| {\cal A} \right|$  in the SM. 
The inclusion of charged Higgs
does not alter this equality on account of Eq.(\ref{eq:ampratio}).
Furthermore, being suppressed by $1/\tan^4{ \beta}$, the charged
Higgs effect on $B\bar{B}$ mixing is negligible if $|\xi| \le 1$.  
Eq.(\ref{eq:atxi}) is therefore applicable to the $CP$ study of 
this quark transition in both $B_d$ and $B_s$ decays to $CP$
eigenstates. 

\vskip .1in
\underline{\bf (i) $B_d$ decays: $B_d \to \psi K_S$}
\vskip .1in

  In the SM, one of the most studied and cleanest observables is the 
$CP$ asymmetry in $B_d \to \psi K_S$. 
As the final state is $CP$-odd ($f_{CP}=-1$),
one has from Eq.(\ref{eq:atxi})
\equation
a(t)_{\rm SM} = - \sin 2\beta_{\rm CKM} \sin (\Delta M t) \, ,
\endequation
where $\beta_{\rm CKM} \equiv \arg\left(-V_{cd}V_{cb}^\ast/
V_{td}V_{tb}^\ast\right)$. The current SM fit gives 
$\sin 2\beta_{\rm CKM} = 0.75 \pm 0.10$ \cite{paganini}.
Recently, the CDF collaboration has reported a preliminary measurement
of the $CP$ asymmetry in $B_d \to \psi K_S$: 
$a_{\psi K_{S}} = 0.79^{+0.41}_{-0.44}$ \cite{sin2beta}. Within the SM,
this simply implies $\sin 2\beta_{\rm CKM} = 0.79^{+0.41}_{-0.44}$.
Though not precise, this measurement can already be used to put
constraints on some models of new physics\cite{nir}.

  In the presence of charged Higgs interactions, the $CP$ asymmetry
no longer measures the CKM angle $\beta_{\rm CKM}$. Instead, it is given by
\equation
a(t)= - \sin 2(\beta_{\rm CKM} + \theta) \sin (\Delta M t) \, ,
\endequation 
and the coefficient $\sin  2(\beta_{\rm CKM} + \theta)$ 
could take values quite different from the SM prediction~\cite{ksw}.
This can lead to a distinct signature for new physics.
As CDF and other experiments improve the measurement of 
$a_{\psi K_S}$, they would put important constraints on the
parameters of the T2HDM.
Indeed, the current CDF measurement \cite{sin2beta} disfavors a large,
negative $\gamma$ for the T2HDM, as can be seen from Fig.~\ref{figcpall}.

\vskip .1in
\underline{\bf (ii) $B_s$ decays: $B_s \to D_s^+ D_s^-$ and 
$B_s \to \psi \eta/\eta^\prime$} 

\vskip .1in

  One of the notable features of the  $B_s\bar{B}_s$ system is that
  the $\Delta M$ mixing rate is much larger than that of its $B_d$ 
counterpart. In the SM, this is  mostly due to the $|V_{ts}/V_{td}|^2$ 
enhancement factor.  Present experimental data give a lower limit 
of \cite{pdg} 
\equation
 \Delta M_s > 9.1  \ {\rm ps}^{-1}  \;\;\; (x_s > 14 ) \, .
\endequation
  Time-integrated $CP$ asymmetry will be strongly suppressed 
by a factor of $x_s/(1+x_s^2)$ and will be extremely difficult to measure
in the near future. Time-dependent $CP$ asymmetry will, on the other 
hand, require excellent time resolution of the detector, and is also
likely to be very difficult.
For definiteness, we will consider time-dependent asymmetry only. 

  Among the exclusive channels available to $B_s$ ($\bar{B}_s$)
decays through $\bar{b} \to c \bar{c} \bar{s}$ ($b \to c \bar{c} s$),
$B_s \to \psi \phi$ has a relatively large branching ratio.
Its final state, however, is not a $CP$ eigenstate 
and $CP$ asymmetry analysis requires information about
the angular distribution of its final state.
  On the other hand, $B_s \to \psi \eta/\eta^\prime$ and 
$B_s \to D_s^+ D_s^-$ decays are quite simple to analyze. 
As for $B_d \to \psi K_S$, Eq.~(\ref{eq:atxi}) is applicable in these
cases.

  It is easily seen that in the Wolfenstein parameterization 
$\phi_D \simeq \phi_M \simeq 0$ in the SM.
Therefore, the SM does not give rise to any $CP$ asymmetry,
\equation 
 a(t)_{\rm SM} \simeq 0 \, .
\endequation
 This singles out $B_s$ decays as unique probes for
new sources of $CP$ violation.

 As noted before, in the T2HDM with $|\xi| \sim 1$, charged Higgs 
contribution to $B_s \bar{B}_s$ mixing is negligible compared to the SM
contribution. The decay phase is simply given by $\theta$.
Since $f_{CP}=1$ for both $B_s \to D_s^+ D_s^-$ and
$B_s \to \psi \eta/\eta^\prime$, one has from Eq.(\ref{eq:atxi}), 
\equation
 a(t) =  \sin (2 \theta ) \sin (\Delta M_s t) \; .
\endequation
The size of the $CP$ asymmetry depends on the CKM phase $\gamma$ 
and the Higgs phase $\delta$, and could be of order tens of percents,
as shown in Figs.~\ref{figcpall}.

\subsection{$b\to c\bar{c} d$}

  Compared to $b\to c\bar{c} s$, the $b\to c\bar{c} d$ transition 
is not as clean. This is due to the fact that the penguin contribution,
though loop-suppressed, suffers no CKM suppression relative to the
tree amplitude.
As a first approximation, we will neglect the pure penguin 
amplitude and take $\left|\overline{\cal A}\right| 
=\left| {\cal A} \right|$. As for $b\to c\bar{c} s$ transition,
this relation remains valid in the presence of charged Higgs exchange
in the factorization approach.

\vskip .1in
\underline{\bf (i) $B_d$ decays: $B_d \to D^+ D^- $}
\vskip .1in
 
  The final state of the decay $B_d \to D^+ D^- $ consists of both
$I=0$ and $I=1$ amplitudes.   
However, to the extent that the pure penguin contribution 
can be neglected,  the decay amplitude has a single
weak phase ${\rm arg}(V_{cb} V_{cd}^*)$, and we can take 
$\left|\overline{\cal A}\right|=\left| {\cal A} \right|$ and use 
Eq.~(\ref{eq:atxi})  to obtain the $CP$ asymmetry,
\equation
 a(t)_{\rm SM} = \sin 2 \beta_{\rm CKM} \sin(\Delta M t) \; ,
\endequation
where we have used $f_{CP}=1$.

\begin{figure}[ht]      
\centerline{\epsfxsize 6.5 truein \epsfbox{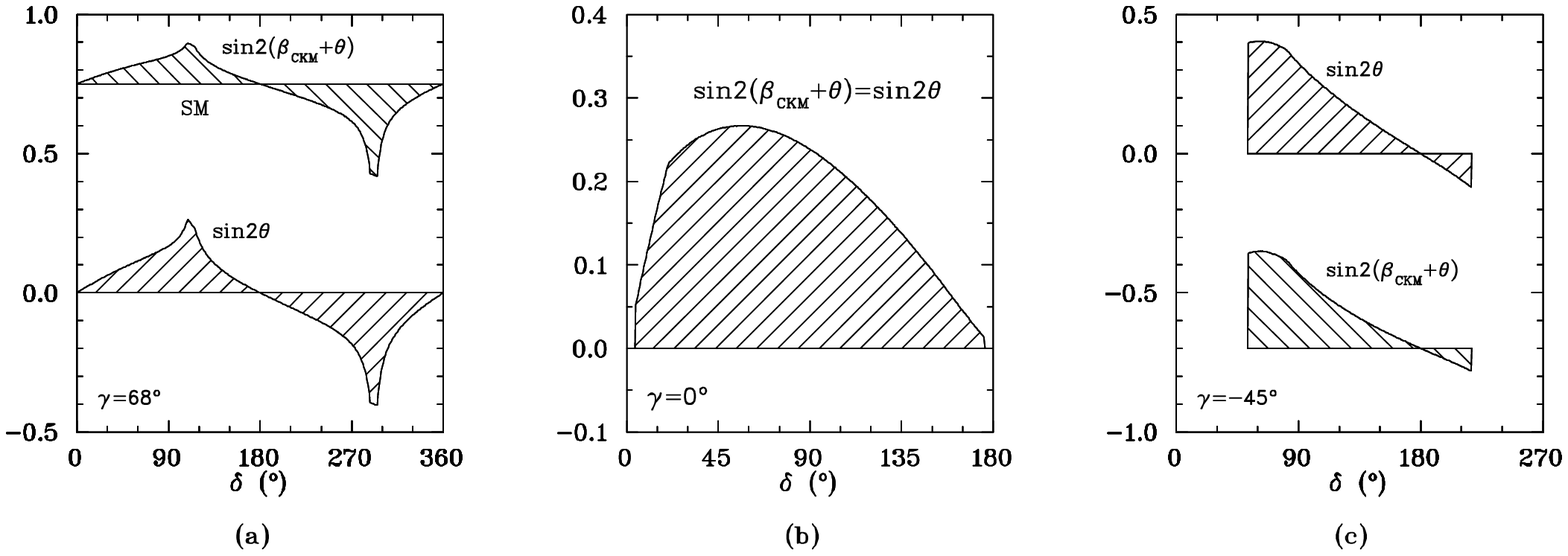}} 
\vskip .2 cm 
\caption[]{
\label{figcpall}
\small  
Allowed regions (shaded) of the $CP$ asymmetries 
$\sin 2 \theta$ and $\sin 2(\beta_{\rm CKM}+\theta)$
in the T2HDM (taking $\xi=e^{-i\delta}$) 
for three representative choices of $\gamma$:
 (a) $\gamma=68^{\circ}$ is a best fit value~\cite{paganini} of the SM,
(b) $\gamma=0^{\circ}$ corresponds to a real CKM matrix,
 and (c) $\gamma=-45^{\circ}$.
The top horizontal line in (a) is for the SM assuming a best fit
$\sin 2\beta_{\rm CKM}=0.75$~\cite{paganini}.
The most stringent constraints from $\Delta m_K$,
$\epsilon_K$, and $b\to s \gamma$ have been imposed.
Note that if the CDF central value for 
$a_{\psi K_S}$ \cite{sin2beta} persists with improved measurements,
 the scenario with a large and negative $\gamma$ (scenario $(c)$)
 will be ruled out. }
\end{figure}

  The inclusion of charged Higgs interactions introduces a new weak phase
in the decay amplitude, and one generally expects direct $CP$ violation to 
occur from interference between the two isospin amplitudes.
However, if the direct $CP$-violating effect is small,
the $CP$ asymmetry then takes the simple form,
\equation
 a(t) = \sin 2 (\beta_{\rm CKM} + \theta)  \sin(\Delta M t) \; ,
\endequation
similar to the decay $B_d \to \psi K_S$.
The amplitude of the asymmetry $\sin 2 (\beta_{\rm CKM} + \theta)$
can take a wide range of values, distinctive from the SM expectation
 (see Fig.~\ref{figcpall}).

\vskip .1in 
\underline{\bf (ii) $B_s$ decays: $B_s \to \psi K_S $}
\vskip .1in

  Compared to the $B_d \to D^+ D^-$ decay, $B_s \to \psi K_S $
is a cleaner mode for $CP$ study for the  following reason.
The final state is a pure $I=1/2$ state with orbital angular momentum
 $l=1$, and the  penguin contribution is expected to be small 
due to both loop and color suppressions. 
As a result,  direct $CP$ violation is suppressed  in the SM.
As for $B_s \to \psi \eta/\eta^\prime$,
 the SM prediction for the $CP$ asymmetry in the
interference between decays with and without mixing is approximately
zero,
\equation
 a(t)_{\rm SM} \simeq  0 \; .
\endequation

   In the T2HDM with $|\xi| \sim 1$, charged Higgs affects only the decay 
but not the mixing amplitude. Similar to the case 
$B_s \to \psi \eta/\eta^\prime$, the $CP$ asymmetry is simply given by
\equation
 a(t) = - \sin 2 \theta  \sin(\Delta M_s t) \; , 
\endequation
where we have used $f_{CP}=-1$.
 The amplitude $\sin 2 \theta$ could be of order tens of percent,
in sharp contrast to the SM prediction.

\subsection{Other $B$-decays}

 In this subsection, we discuss neutral $B$ decays 
involving the $u$ quark Yukawa couplings $H^+\bar{u}_Rq_L$ ($q=d,s,b$).
These couplings depend on $\xi^{\prime}$,
the mixing parameter between the first and the third families in $U_R$.
The size of $\xi^{\prime}$ is constrained by the
$B \to \tau \nu$ rate as given by Eq.~(\ref{eq:Btaunubound}). 
  There are six charged-current four-quark operators for $b$ decays
involving the $u$ quark,
$b\to c \bar{u} d$, $b\to u \bar{c} d$,
$b\to c \bar{u} s$, $b\to u \bar{c} s$,
$b\to u \bar{u} d$, $b\to u \bar{u} s$.
In the T2HDM, only  $b\to u \bar{c} s$, $b\to u \bar{c} d$,
and $b\to u \bar{u} s$ 
could receive sizable charged-Higgs contributions 
compared to their corresponding SM amplitudes.
The phenomenological implications can be summarized as follows:

\begin{itemize}

\item $b \to u\bar{u}d$ and $b \to d\bar{d}d$:
  The charged Higgs effect is small after imposing the $\xi^\prime$
constraint.  The decay  $B_d \to \pi \pi$ remains as in the SM, and 
one may use isospin analysis \cite{GL} to extract the UT angle
$\alpha$.

\item $b \to u\bar{u}s$ and $b \to d\bar{d}s$:
Here, the SM amplitudes are dominated by penguin contributions
\cite{cleopenguin}. In the T2HDM, even though tree-level Higgs-change
effect is negligible, charged-Higgs-mediated penguin contributions
could become appreciable.
Consequently, the procedure to extract the angle $\gamma$ from
the $B^\pm \to \pi K, \; \pi \pi$ decays \cite{Kpi} may be modified.

\item $b \to c \bar{u} d$ and $b\to u \bar{c} d$:
  Both decays are free from penguin contributions.
In the SM, the latter amplitude is Cabibbo-suppressed by 
${\cal O}(\lambda^2)$ with respect to the former, and
 $CP$ asymmetry measurements in the hadronic decays
$B_d \to D_{CP} \pi$ and $B_d \to D_{CP} \rho$ provide a clean
way to extract the angle $\beta_{\rm CKM}$. On the other hand,  the decay
$B_s \to D_{CP} K_S$ is expected in the SM to have an approximately 
zero $CP$ asymmetry.
  Charged-Higgs exchange in the T2HDM has a negligible effect on 
$b \to c \bar{u} d$, whereas its contribution to the Cabibbo-suppressed decay
$b\to u \bar{c} d$ can be  sizable but still smaller than the SM one.
As a result, the $CP$ asymmetries remain the same as the SM
prediction.

\item $b \to u\bar{c}s$ and $b \to c\bar{u}s$:
These transitions can mediate the charged $B^\pm \to K^\pm D/\bar{D}$ decays,
which may provide a clean way within the SM to extract the angle 
$\gamma$ \cite{ADS}. Charged-Higgs-exchange contribution in the T2HDM
will be discussed in detail in the next section on direct $CP$ violation.

\end{itemize}

   Some main results of our analysis concerning $B_d$ and $B_s$ decays
are summarized in Table~2.

\newpage
Table 2:  $CP$ asymmetries for neutral $B$ decays to $CP$ eigenstates
in the SM and in the T2HDM. For $B_s \to \rho K_S$, 
the pure-penguin contribution may be important~\cite{cleopenguin}.
As a result, the tree- and  pure-penguin-amplitudes of
different weak (and possibly strong)  phases are competitive, 
and the $CP$ asymmetry may not be simply related to the CKM phase.
The overall sign of the asymmetry in $B_d \to D_{CP} \rho$ depends
on the $CP$ properties of $D_{CP}$.
\begin{center}
\begin{tabular}{|c|c|c|c|c|c|c|}
\hline \hline
quark  & \multicolumn{3}{c|}{$B_d$ decays} & 
               \multicolumn{3}{c|}{$B_s$ decays} 
\\ \cline{2-7}
  transitions & final states & $\;\;\;\;$ SM $\;\;\;\;$  & 
   $\;\;\;\;\;$ T2HDM $\;\;\;\;\;$
               & final states &$\;\;\;$ SM $\;\;\;$ &$\;\;$ T2HDM $\;\;$
\\ \hline
$b \to c \bar{c} s$ & 
  $\psi K_S$ & $-\sin 2 \beta_{\rm CKM}$ & $-\sin 2(\beta_{\rm CKM} + \theta)$
 & $D_s^+ D_s^-$, $\psi \eta/\eta^\prime$ &  0 & $\sin 2 \theta$ 
\\ \hline
$b \to c \bar{c} d$ &
 $D^+ D^-$ & $\sin 2 \beta_{\rm CKM}$ & $\sin 2(\beta_{\rm CKM} + \theta)$
& $\psi K_S$ &  0 & $-\sin 2 \theta$ 
\\ \hline
$b \to c \bar{u} d$ & $D_{CP}\rho$ &  $\pm \sin 2 \beta_{\rm CKM}$ &  
$\pm \sin 2 \beta_{\rm CKM}$ &
                      $D_{CP} K_S$ & 0 & 0
\\ \hline
$b \to u \bar{u} d$ & $\pi \pi$ & $- \sin 2 \alpha$ &  $- \sin 2 \alpha$ &
                       $\rho K_S$ & competing & competing
\\ \hline \hline
\end{tabular}
\end{center}
\vskip 0.1in

\section{Direct CP violation in $B$ decays}

\subsection{Direct CP violation in $B^{\pm} \to \psi/J K^{\pm} $ }

 In this subsection, we address the use of charged 
$B$ decays to search for direct CP violation.
This requires a difference in the strong phases associated with the $W$- and
$H$-mediated decay amplitudes leading to the exclusive final
state(s) of interest.
As an illustration, 
we will focus our attention on the decay
$B^{\pm} \to \psi/J K^{\pm} $.  There are several features about this
mode which make it interesting to study both experimentally and
theoretically.  
The $B^{\pm} \to \psi/J K^{\pm} $ decays proceed through
the quark decay $b \to c \bar{c} s$ and its conjugate process.
In the SM, the weak phase associated with the decay amplitude
is vanishingly small;
therefore,  the SM predicts a zero rate asymmetry.
Second, it is clean to measure experimentally.
Third, it has a large branching ratio of $\sim 10^{-3}$ \cite{pdg}.
These features make the experimental measurement very
worthwhile in the search for new sources of $CP$ violation.
However, as we can not reliably compute the hadronic matrix elements of 
either the SM current-current four-Fermi
operator or  the charged-Higgs-induced scalar operator,
the relative strong phase between the $W$- and $H$-mediated amplitudes
remains largely unknown.
Therefore, no reliable predictions can be made about the size of the
asymmetry.

Formally, the amplitude for $B^+ \to \psi/J K^+$ can be written as
\equation
 {\cal A} = {\cal A}_{SM} [ 1 - \zeta e^{-i\delta} e^{-i\phi_s} ].
\endequation
where $\delta$ and $\phi_s$ are the weak and strong phase difference
respectively.
It then follows that the amplitude for $B^- \to \psi/J K^-$ is given as
\equation
 \overline{\cal A} = \overline{\cal A}_{SM} 
[ 1 - \zeta e^{+i\delta} e^{-i\phi_s} ].
\endequation

\begin{figure}[ht]      
\centerline{\epsfxsize 4 truein \epsfbox{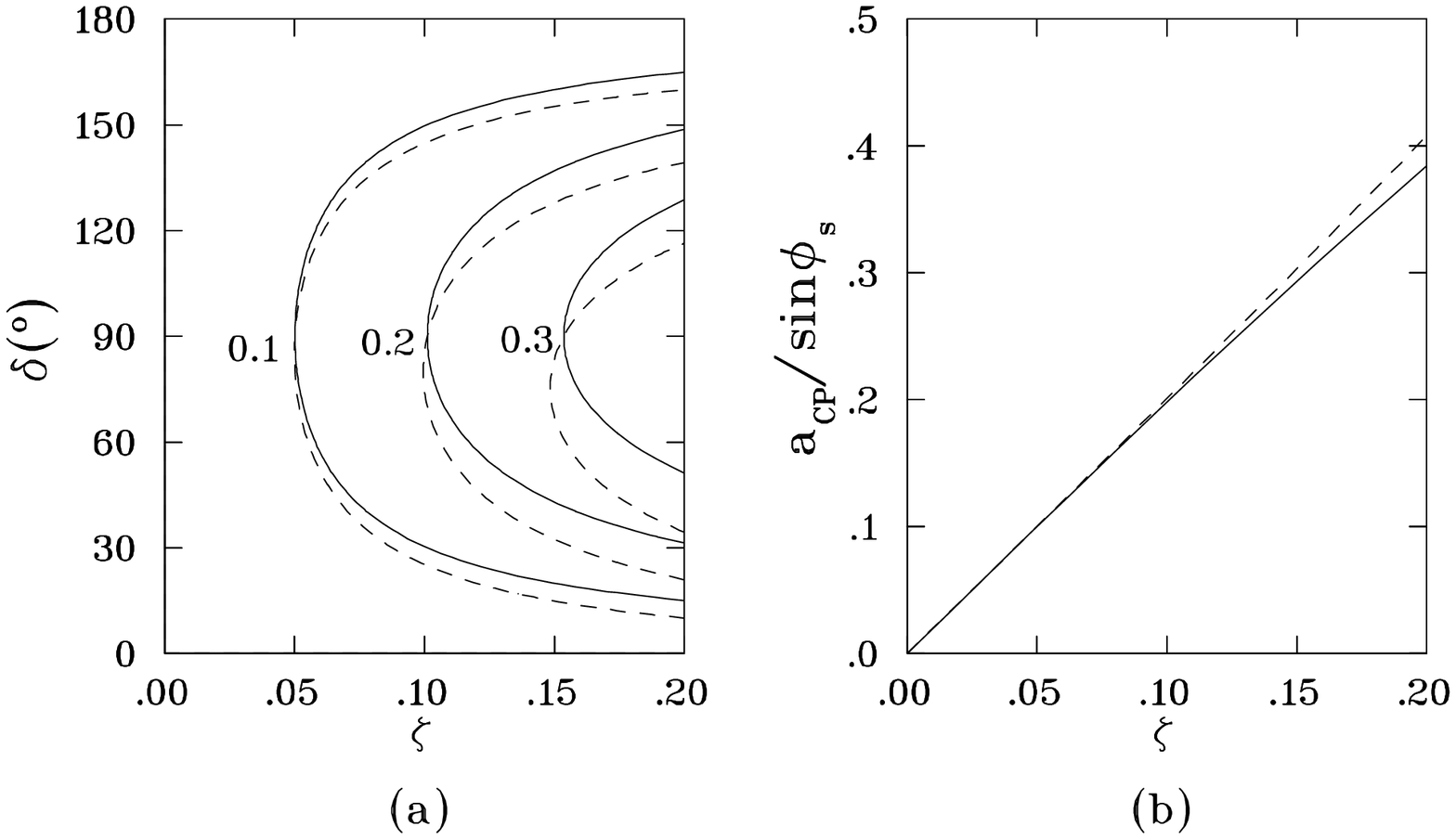}} 
\vskip .2 cm 
\caption[]{
\label{figdcp}
\small  Direct CP asymmetry with two representative values
for the strong phase: $\phi_s=90^{\circ}$ (solid lines)
and $\phi_s=30^{\circ}$ (broken lines). (a) is the contour plot
for $a_{CP}/\sin \phi_s$ with contour levels 0.1, 0.2, and 0.3
respectively, (b) is the maximum value $a_{CP}/\sin \phi_s$ takes
as a function of $\zeta$. }
\end{figure}

The CP-violating partial rate asymmetry (PRA) can be expressed as
\equation
 a_{CP} = \frac{\Gamma(B^+ \rightarrow K^+ \psi) - \Gamma(B^- \rightarrow
K^- \psi)}{\Gamma(B^+ \rightarrow K^+ \psi) + \Gamma(B^- \rightarrow
 K^- \psi)}
  = \frac{2 \zeta \sin \delta \sin \phi_s }
         { 1+ \zeta^2 - 2 \zeta \cos \delta \cos \phi_s}  \; .
\endequation
As $a_{CP}$ is directly  proportional to $\zeta$ when $\zeta$ is small,
it is crucial to determine how large $\zeta$ can be after imposing
the experimental constraints. 
From section~\ref{sec:constraints}, we know that the most stringent
constraints on $\zeta$ come from the $b\to s \gamma$ rate, 
the $K_L$-$K_S$ mass difference $\Delta m_K$, and $\epsilon_K$.
For $|\xi|=1$, the combined constraints from $\Delta m_K$ 
(see Eq.(\ref{eq:mkbound}))
and $b\to s \gamma$ (see Fig.~\ref{figbsg}) 
imply an upper bound of
\begin{eqnarray}
\zeta  & \le  & 0.2 \; .
\end{eqnarray}
As can be seen from Fig.~\ref{figbsg},
this limit is saturated for $\delta \le 45^\circ$, and becomes more
stringent
as the phase $\delta$ increases all the way to $180^\circ$.  
Imposing the $\epsilon_K$ constraint may generally make $\zeta$ even 
smaller, depending on the CKM phase and $\delta$ (see Fig.~1 of 
Ref.\cite{ksw} for an illustration).
Suffice it to say that after taking all the data into account,
$a_{CP}$ could be of order $10\%$ in the T2HDM
unless the strong phase $\phi_s$ is much suppressed, as 
is illustrated in Fig.~\ref{figdcp}.

  Recall that in the SM, the PRA has to come from the interference
between the tree diagram and a loop diagram with internal $u$-quark which
gives the required weak phase difference, 
Arg[$V_{ub}V^*_{us}/V_{cb}V^*_{cs}$]. 
Consequently, the rate asymmetry in the SM is both loop- and 
Cabibbo-suppressed and is expected to be vanishingly small.
In the T2HDM, on the other hand, the $CP$ asymmetry arises from the
interference between two tree diagrams, and may only be  suppressed
by the relative strength of the two amplitudes, $\zeta$.
As a result, rate asymmetry at the $10\%$ level can be naturally 
expected with $\zeta \sim 0.1$, at least in those exclusive channels
where the strong phase difference is not too suppressed.

We remark, in passing, that the CPT theorem does not impose 
any particular restriction on a specific channel (such as
$\psi/J K$ ), materializing from $c \bar{c} s$  
as there are many other channels originating from the 
same quark transition.
Indeed, in the next subsection, we will discuss a similar 
case where large direct $CP$ asymmetry is expected,
based purely on the SM, in exclusive channels emerging
from quark-level interference between two tree graphs.

To estimate the number of $B$'s needed  in order to
measure a  $10\%$ asymmetry at the $2\sigma$ level,
we assume the detector efficiency to be $25\%$.
Then for the needed number of $B$'s we find
$\frac{2^2}{10^{-3}\times 0.12 \times 0.1^2 \times 0.25}
\simeq 1.3 \times 10^7$.
In deducing this number, we have used $BR(\psi/J \to e^+e^-, \mu^+ \mu^-)
=12\%$. This estimate suggests that such an analysis 
may be doable at various $B$ facilities.
Indeed, the existing data sample  ($\sim 10^7 \; B$'s) at CLEO
can already be used for this important search\cite{bcp3}.

 It is worth pointing out that the possibility of sizable direct CP violating 
effect in $B^\pm \to \psi/J K^\pm$ in the T2HDM 
does not arise in 2HDM's with NFC (Models I and II) where
the $W$ and $H$ exchange amplitudes have the same weak phase.
Note also that large CP asymmetries may also be expected in 
charged $B$ decays 
through $b \to c \bar{c} d$ and its conjugate process, for example,
in $B^{\pm} \to \psi/J \pi^{\pm}$ 
(the SM expectation for the $CP$ asymmetry in this mode is small).
However, the branching ratio for this decay is Cabibbo-suppressed by
a factor of $20$ compared to that of  $B^\pm \to \psi/J K^\pm$.

Before leaving this subsection, we want to emphasize 
that there are  many exclusive channels available 
through the $\bar{b} \to c \bar{c} \bar{s}$ transition, {\it e.g.} 
$\psi K^+$, $\psi^\prime K^+$, $\psi K^+ \pi \pi$,
$\bar{D} D^+_s$, $D \bar{D} K^+$, $\cdots$, 
that may offer ample opportunities 
for sizable strong phase difference ($\phi_s$) for certain modes.
It is thus important that experimental searches for direct $CP$ violation
in charged $B$ decays include as many of these modes as possible
and not be restricted to the  $B^\pm \to \psi/J K^\pm$ 
channel only.

\subsection{CP violation in $B^\pm \to D/\bar{D} K^\pm$ 
           and the extraction of $\gamma$ }

  The decays $B^+ \to  K^+ D$ and  $B^+ \to  K^+ \bar{D}$ proceed 
respectively through the transitions $\bar{b} \to \bar{u} c \bar{s} $ and 
$\bar{b} \to \bar{c} u \bar{s}$.
Although the decay amplitude for $B^+ \to K^+ D$ 
is both color- and Cabibbo-suppressed relative to $B^+ \to K^+ \bar{D}$, 
the amplitudes for decays to certain common final states via
$B^+ \to  K^+ D [ \to f]$ and $B^+ \to K^+ \bar{D} [\to f]$
can be expected to be comparable. This is due to the fact that 
$D$ decays into $f$ through the Cabibbo-allowed
channel $c \to s \bar{d}u$ and $\bar{D}$ decays 
through the doubly-Cabibbo-suppressed channel 
$\bar{c} \to \bar{d} s \bar{u}$ 
into the same final state $f$. 
Examples of such common final states of the $D/\bar{D}$ decay
include $f=K^- \pi^+, K^- \pi^+ \pi^0, K^-\rho^+, K^- a_1^+, K^{*-} \pi^+,$
 etc. 
Atwood, Dunietz, and Soni (ADS) \cite{ADS} observed that within the SM 
the large interference effect between the two decay chains
may give rise to  ${\cal O}(1)$ partial rate asymmetries 
(PRA) between the $B^+$ and $B^-$ decays, 
\begin{eqnarray}
a_{Kf} & \equiv & 
 \frac{\Gamma(B^+ \to K^+ [f]_D) - \Gamma(B^- \to K^- [\bar{f}]_D)}
      {\Gamma(B^+ \to K^+ [f]_D) + \Gamma(B^- \to K^- [\bar{f}]_D)} \; .
\end{eqnarray}
Furthermore, Ref.\cite{ADS} suggests that the angle $\gamma$ of
the unitarity triangle can be extracted from a study involving 
a minimum of two different final states $f$.
In this section, we examine the charged Higgs effect on the direct
CP asymmetry and on the extraction of $\gamma$ using the
$B^\pm \to D/\bar{D} K^\pm$ decays.

  In the T2HDM, due to the small Yukawa couplings,
$H^+\bar{u}(1 \pm \gamma_5) q$ ($q=d,s$),
 the charged Higgs amplitudes for 
$c \to s \bar{d}u $  and $\bar{c} \to \bar{d} s \bar{u}$ 
are negligible compared to the corresponding SM amplitudes. 
The hadronic decays $D (\bar{D}) \to f$ thus remain as in the SM. 
Because of the same suppression, the Higgs contribution to 
$B^+ \to K^+ \bar{D}$ 
is at most a few percent of the SM amplitude and can also be safely neglected. 
On the other hand, the charged Higgs amplitude for the color- and 
Cabibbo-suppressed decay $B^+ \to K^+ D$ could in principle be sizable as it 
involves
the potentially large Yukawa's $H^+\bar{u}_R b_L$ and $H^+\bar{c}_R s_L$. 
However, the present experimental limit on the $B^+ \to \tau^+ \nu$ rate 
already places an upper bound
of about $25\%$ on the decay-amplitude ratio $|{\cal A}_H/{\cal A}_W|$.
For simplicity of the analysis, we will first ignore the charged-Higgs
 contribution to the decay  $B^+ \to K^+ D$.

  In the SM, $D\bar{D}$ mixing is vanishingly small.
 As discussed in Sect. \ref{sect:DD}, however, 
the $D\bar{D}$ mixing parameter, 
$x_D\equiv \Delta m_D/\Gamma_D$, could be as large as 
a few percent in the T2HDM.
This will in turn affect the direct $CP$ asymmetry in
$B^+ \to K^+ [f]_D$ and  $B^- \to K^- [\bar{f}]_D$ decays,
as well as the standard procedure to extract $\gamma$ from
these decays.
  Note that in 2HDM's with NFC (Models I and II), charged Higgs
effect on $D\bar{D}$ mixing and on the relevant $B$ and $D$ decays
is small, and the ADS method \cite{ADS} to extract $\gamma$ remains unmodified.
We now analyze in the T2HDM this new effect due to $D\bar{D}$ mixing.
  
  Recall that in the absence of $D\bar{D}$ mixing, as in the SM,
the total decay amplitude for $B^+$ can be written as \cite{ADS}
\begin{eqnarray}
{\cal A}_{ADS} & = & {\cal A}_{B^+ \to K^+ D} {\cal A}_{D \to f}
             +{\cal A}_{B^+ \to K^+ \bar{D}} {\cal A}_{\bar{D} \to f} \; .
\end{eqnarray}
In the SM, we can parameterize 
${\cal A}_{B^+ \to K^+ D}/{\cal A}_{B^+ \to K^+ \bar{D}}=
 r_B e^{i\gamma} e^{i \Delta_B}$, and
${\cal A}_{\bar{D} \to f}/{\cal A}_{D \to f} =
 - r_D  e^{i \Delta_D}$, with $\Delta_B$ and  $\Delta_D$  denoting 
the strong phase differences in the $B$ and $D$ decay amplitudes respectively.
The amplitude ratios can be estimated up to uncertainties in the
hadronic matrix elements,
$r_D \sim |(V_{cd}V_{us}^*)/(V_{cs}^* V_{ud})| \sim 0.05$ for $D$ decays, 
 and $r_B \sim |(V_{ub}^*V_{cs})/(V_{cb}^*V_{us})||a_2/a_1| \sim 0.08$,
where we have used $|V_{ub}/V_{cb}| \sim 0.08$ and $|a_2/a_1|\sim 0.21$
\cite{a2a1}
accounts for the color suppression factor. 
As we will include the effect due to $D\bar{D}$ mixing,
we cannot use the $D$ and $\bar{D}$ decay branching ratios to extract
$r_D$.  
The partial rate asymmetry can be easily obtained as,
\begin{eqnarray}
a_{Kf,SM} & = & 
\frac{2r_Br_D \sin \gamma \sin (\Delta_B - \Delta_D)}
   { r_B^2 + r_D^2 -2 r_B r_D \cos \gamma \cos (\Delta_B - \Delta_D)} \; .
\end{eqnarray}
Note that a non-zero strong phase difference is required for the
asymmetry and for the extraction of $\gamma$, and the asymmetry
need not be small.

  Current data set an upper limit on the $D\bar{D}$ mixing parameter
$x_D < 0.1$, and this allows us to approximate the $D\bar{D}$ mixing
effect by keeping terms up to the quadratic order in $x_D$.
$D\bar{D}$ mixing thus leads to one additional contribution 
to the $B$ decay amplitude:
$B^+$ decays through the color-allowed channel
$B^+ \to K^+ \bar{D}$, followed by the subsequent oscillation $\bar{D}\to D$
with $D$ decaying to the Cabibbo-allowed final state, 
e.g. $f=K^-\pi^+$.
Neglecting the other $D\bar{D}$-mixing-induced decay path 
which involves both the color-suppressed $B$ decay and the 
doubly-Cabibbo-suppressed $D$ decay,
this new contribution can be written as \cite{MecaSilva}
\begin{eqnarray}
{\cal A}_{D\bar{D}} & = & 
{\cal A}_{B^+ \to K^+ \bar{D}} \frac{p_D}{q_D} g_-(t)
 {\cal A}_{D \to f} \; .
\end{eqnarray}
Here up to quadratic order in $x_D$, 
$g_-(t) \simeq \frac{i}{2} x_D \tau_D e^{-\tau_D/2}$ with 
$\tau_D = \Gamma_D t$,  
 and $\frac{p_D}{q_D}=e^{-i2\theta_D}$ with the $D\bar{D}$ mixing phase
$\theta_D={\rm arg}(\xi) - {\rm arg}(\xi^\prime)$ in the T2HDM.
 
  The time-integrated $B$ decay width including the $D\bar{D}$ mixing effect
is then given by 
\begin{eqnarray}
\Gamma(B^+ \to K^+ [f]_D) &  \sim  &
r_B^2 + r_D^2  -2 r_B r_D \cos(\gamma + \Delta_B - \Delta_D)
\nonumber \\
 &&     + x_D r_B \sin(\gamma + 2 \theta_D + \Delta_B)
      - x_D r_D \sin (2\theta_D + \Delta_D)
      + x_D^2/2
\\
\Gamma(B^- \to K^- [\bar{f}]_D) & \sim & 
r_B^2 + r_D^2  -2 r_B r_D \cos( \Delta_B - \Delta_D -\gamma) 
\nonumber \\
  &&    + x_D r_B \sin(\Delta_B - \gamma - 2 \theta_D)
      - x_D r_D \sin (\Delta_D - 2\theta_D )
      +  x_D^2/2 \; ,
\end{eqnarray}
where the $x_D r_B$  and $x_D r_D$ terms come from the interferences 
between ${\cal A}_{D\bar{D}}$ and the color-suppressed $B$-decay  amplitude 
and the doubly-Cabibbo-suppressed $D$-decay  amplitude respectively.
The $CP$ asymmetry can then  be obtained as,
\begin{eqnarray}
a_{Kf} & = & \frac{2r_Br_D \sin\gamma \sin(\Delta_B -\Delta_D)
           + x_D r_B \sin(\gamma + 2\theta_D) \cos\Delta_B
           - x_D r_D \sin 2\theta_D \cos \Delta_D }
    {r_B^2 + r_D^2 
     - 2 r_B r_D \cos \gamma \cos(\Delta_B -\Delta_D)
     + \Gamma_{x_D} }
\end{eqnarray}
where $\Gamma_{x_D} =
    x_D r_B \cos(\gamma + 2\theta_D) \sin\Delta_B
     - x_D r_D \cos2\theta_D\sin \Delta_D 
   + x_D^2/2 $.
  
\begin{figure}[ht]      
\centerline{\epsfxsize 4 truein \epsfbox{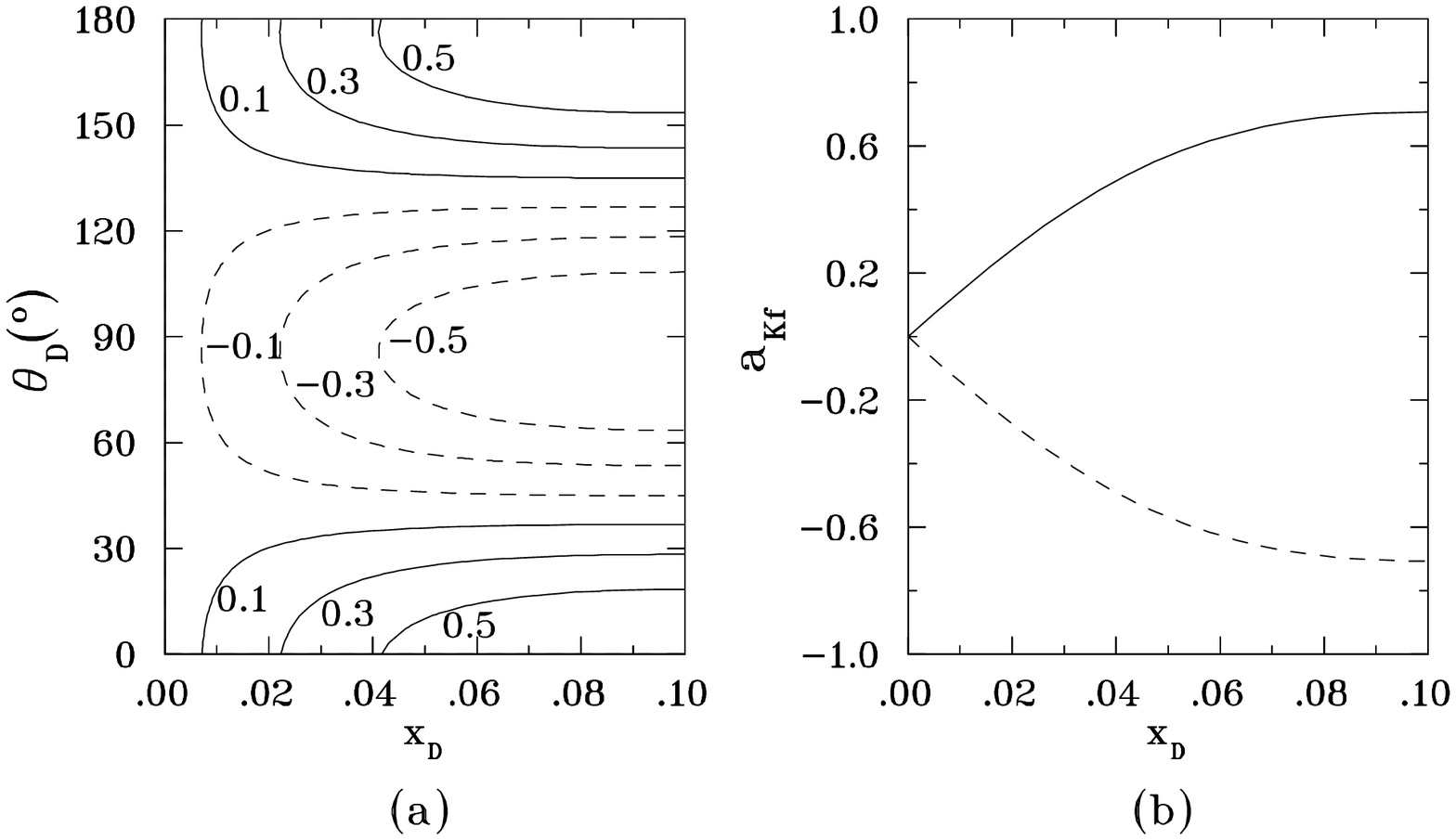}} 
\vskip .2 cm 
\caption[]{
\label{figdKfB0D0}
\small  $D\bar{D}$-mixing-induced direct $CP$ asymmetry $a_{Kf}$
as given by Eq.~(\ref{eq:axd}), assuming the strong phase differences
$\Delta_B$ and $\Delta_D$ to be zero.
We have taken $r_B=0.08$, $r_D=0.05$, and $\gamma=60^\circ$ for the
numerical analysis.
Shown in (a) is the contour plot
for $a_{Kf}$ with contour levels 0.1, 0.3, 0.5 (solid lines)
and -0.1, -0.3, -0.5 (dashed lines).
The solid and dashed lines in (b) denote respectively the maximum and the
minimum values which $a_{Kf}$ can take as a function of $x_D$. }
\end{figure}

  It is interesting to note that whereas a
sizable strong phase difference is essential for a large
rate asymmetry in the SM,
this is no longer necessary with a non-zero $x_D$.
While in general there is no reason for both strong phases
($\Delta_B$ and $\Delta_D$) to be small,
for illustration, let us consider the $CP$ asymmetry in the limit 
that they are, i.e., for simplicity, we set $\Delta_B=\Delta_D=0$,
which may be a good approximation for certain decay channels.
Then
\begin{eqnarray} \label{eq:axd}
a_{Kf} & \longrightarrow & \frac{x_D r_B \sin(\gamma + 2\theta_D)
                    -x_D r_D \sin 2\theta_D}
            {r_B^2 + r_D^2 - 2 r_B r_D \cos \gamma + x_D^2/2} \; .
\end{eqnarray}
This is the asymmetry due to $D\bar{D}$ mixing.
For $x_D$ of a few percent, $a_{Kf}$ could be of order tens of percent,
comparable to the expected asymmetry within the SM.  
In other words, the presence of $D\bar{D}$ mixing could affect the 
ADS method to extract $\gamma$ in a significant way.
This  $D\bar{D}$ mixing-induced asymmetry $a_{Kf}$
is shown in Fig.~\ref{figdKfB0D0}.

\begin{figure}[ht]      
\centerline{\epsfxsize 4 truein \epsfbox{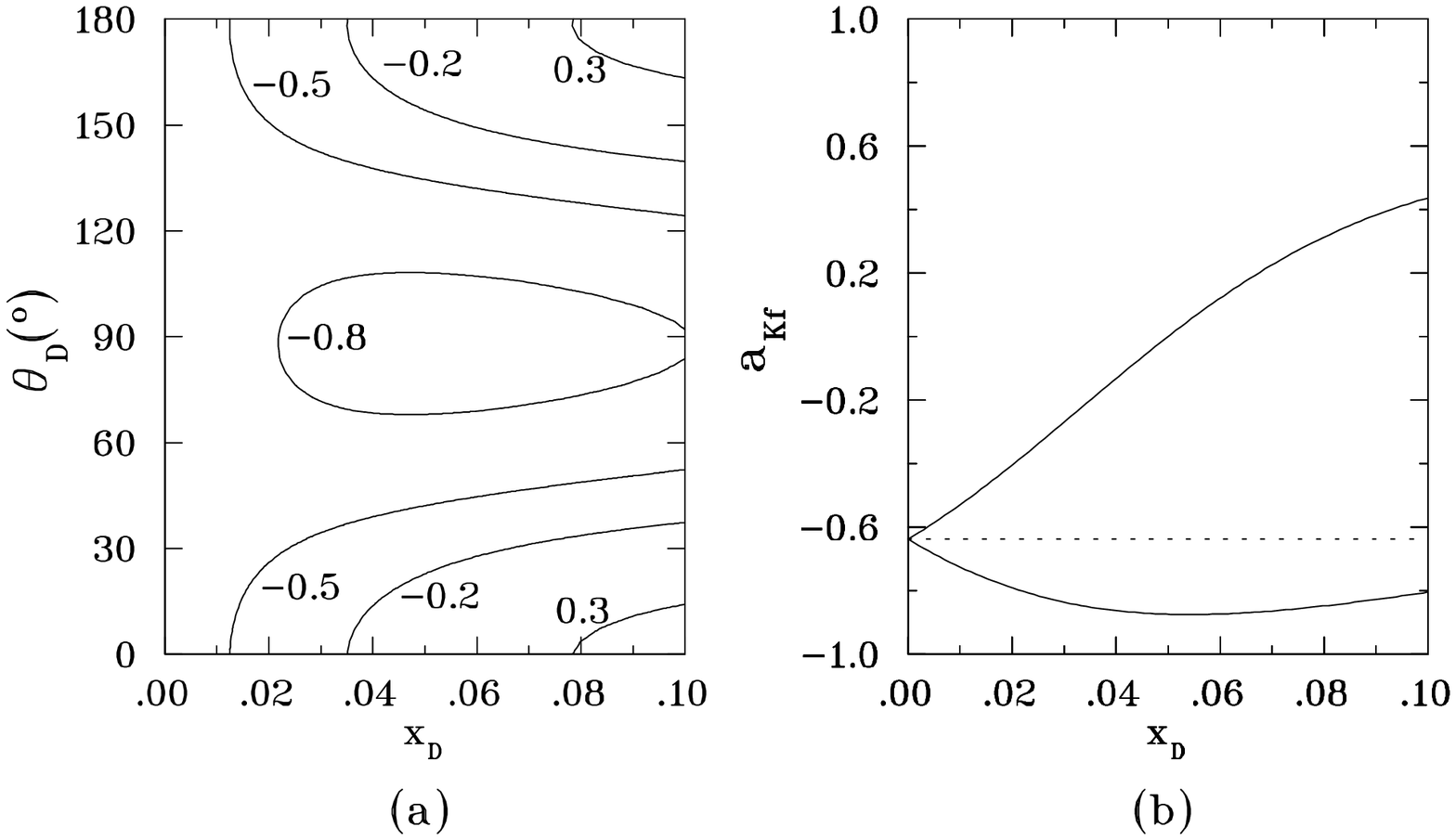}} 
\vskip .2 cm 
\caption[]{
\label{figdKfB0D30}
\small  The effect of $D\bar{D}$ mixing on the PRA $a_{Kf}$ 
for  $\Delta_B=0^\circ$ and $\Delta_D=30^\circ$. 
 We have used $r_B=0.08$, $r_D=0.05$, and $\gamma=60^\circ$ for 
the numerical analysis.
Shown in (a) is the contour plot
for $a_{Kf}$ with contour levels $-0.8$, $-0.5$, $-0.2$, and $0.3$.
The two solid lines in (b) represent respectively the maximum and the
minimum values which $a_{Kf}$ can take as a function of $x_D$.
The broken line in (b) shows the value of $a_{Kf}$ in the absence
of $D\bar{D}$ mixing as expected in the SM. }
\end{figure}

   Given the many decay channels available for both $D$ and $\bar{D}$,
it is highly unlikely that $\Delta_D$ should be small for all the modes.
For illustrative purpose, we also consider the case of
$\Delta_B=0^\circ$ and $\Delta_D=30^\circ$, and 
evaluate the effect of $D\bar{D}$ mixing on the PRA, $a_{Kf}$.
The numerical results are presented in Fig.~\ref{figdKfB0D30}.
One again observes a sizeable effect due to $D\bar{D}$ mixing.

   Figs.~\ref{figdKfB0D0} and \ref{figdKfB0D30} illustrate the effect of
$D\bar{D}$ mixing on two classes of decays with respectively
small and large strong phase differences.
For $x_D$ of order a few percent, one arrives at a no-lose theorem that large
$CP$ asymmetries are expected for both cases.
However, it seems that the angle $\gamma$ can not be extracted in a simple
manner from such decays in the presence of a sizeable $D\bar{D}$ mixing.

\section{Conclusion}

 As has been shown in this work, the phenomenology of the T2HDM differs 
from that of Models I and II 2HDM  (and the SM) in many interesting ways.
These include their different effects on
$K\bar{K}$, $D\bar{D}$, and $B\bar{B}$ mixings, on charged-current decays,
 and on  $CP$-violation in both neutral and charged $B$ decays.
We have examined the $CP$-violating phenomenology associated with
the anomalous charged-Higgs Yukawa couplings in the T2HDM.
The effects of the two new $CP$-violation parameters of the model,
$\xi$ and $\xi^\prime$, nicely separate, with the latter mainly
affecting processes that involve $D\bar{D}$ mixing.
As a result,  some clean predictions for neutral $B$ decays
can be made within the T2HDM as presented in Table~2.
 The $\xi$-related $CP$ angle $\theta$ 
can be directly measured from the
$B_s \to D^+_s D^-_s, \; \psi \eta/\eta^\prime,$ or $\psi K_S$ decay
without SM pollution.
On the other hand, the $\alpha$ and $\beta_{\rm CKM}$ angles of the unitarity 
triangle may be extracted from $CP$ asymmetry measurements in
the decays $B_d\to \pi\pi$ and $B_d\to D_{CP} \rho/\pi$ respectively without
charged-Higgs contamination.
  Based on the information on $\theta$ and $\beta_{\rm CKM}$, 
a cross check on the model can then be made by 
taking into account the $CP$ asymmetry measurement in the 
decays  $B_d \to \psi K_S$ and/or $B_d \to D^+ D^-$.

  Besides $CP$ asymmetries in neutral $B$ decays, the model 
also gives rise to generically large $CP$-violating effects in 
charged $B$ decays. 
For example, the partial rate asymmetry in $B^\pm \to K^\pm \psi$
can be of order ten percent if the strong phase difference
 is not too much  suppressed.
The model can also lead to a sizable $D\bar{D}$ mixing effect, 
depending on the mixing in the right-handed up quark sector.
This in turn could strongly modify the direct $CP$ asymmetry
in $B^\pm \to D/\bar{D} K^\pm$, and thus the procedure to extract the 
CKM angle $\gamma$ from this process.
On the other hand, as in most charged-Higgs models of $CP$ violation,
the direct $CP$-asymmetry in the radiative decay $b\to s \gamma$
is found to be tiny\cite{prep}.

\vskip 0.3in
We thank Nilendra Deshpande and  Ken Kiers for discussions. 
This research was supported in part by the U.S. Department of Energy 
contract numbers  DE-AC02-98CH10886 (BNL) and
DE-FG03-96ER40969 (Oregon).
  
\vskip 0.3in
Note Added in Proof:

 Since this manuscript was sent for publication, the CLEO Collaboration
has published results on their first search  for direct $CP$ violation in
$B^\pm \to K^\pm \psi/J$ 
[See G. Bonvicini {\it et al.},  Phys. Rev. Lett. {\bf 84}, 5940 (2000).].


\begin{thebibliography}{99}
\addtolength{\baselineskip}{-.1\baselineskip}

\bibitem{BaBar}
For recent  reviews, see, for example, 
{\it The BaBar Physics Book}, {\it ed.} P. Harrison and H. Quinn,
SLAC-R-504, 1998; Y. Nir, hep-ph/9911321.

\bibitem{ksw}
 K. Kiers, A. Soni, and G.-H. Wu, Phys. Rev. D {\bf 59}, 096001 (1999);
hep-ph/9903343, talk given at American Physical Society (APS) Meeting 
of the Division of Particles and Fields (DPF 99), Los Angeles, CA, 
5-9 Jan 1999. 
%
\bibitem{daskao}
	A. Das and C. Kao, Phys. Lett. B {\bf 372}, 106 (1996).

\bibitem{type3}
T.P. Cheng and M. Sher, Phys. Rev. D {\bf 35}, 3484 (1987);
W.-S. Hou, Phys. Lett. B {\bf 296}, 179 (1992);
M. Luke and M. Savage, Phys. Lett. B {\bf 307}, 387 (1993);
L. Hall and S. Weinberg, Phys. Rev. D {\bf 48}, R979 (1993);
Y. L. Wu and L. Wolfenstein, Phys.  Rev.  Lett.  {\bf 73}, 1762 (1994),
and Y. L. Wu, hep-ph/9404241; 
D. Atwood, L. Reina, and A. Soni, Phys.\ Rev.\ D {\bf 55}, 3156 (1997);
D. Bowser-Chao, K. Cheung, and W.-Y. Keung, Phys.\ Rev.\ D {\bf 59},
115006 (1999).  

\bibitem{NFC}
S. Glashow and S. Weinberg, Phys. Rev. D {\bf 15}, 1958 (1977).

\bibitem{hhunter}
For a review, see J. Gunion {\it et. al.}, {\em The Higgs Hunter's Guide},
Addison-Wesley Publishing Company, 1990.

\bibitem{ADS}
D. Atwood, I. Dunietz, and  A. Soni, Phys. Rev. Lett. {\bf 78}, 3257 (1997).

\bibitem{topCond}
W.A. Bardeen, C. Hill, and M. Lindner, Phys. Rev. D {\bf 41}, 1647 (1990);
V.A. Miransky, M. Tanabashi, K. Yamawaki, Phys. Lett. B {\bf 221}, 177 (1989).

\bibitem{topColor}
C. Hill, Phys. Lett. B {\bf 266}, 419  (1991); 
{\it ibid.} {\bf 345}, 483 (1995).

\bibitem{heyuan}
H.-J. He and C.P. Yuan, Phys. Rev. Lett. {\bf 83}, 28 (1999).

\bibitem{paganini}
  P. Paganini {\it et. al.}, Phys. Scripta {\bf 58}, 556 (1998); 
  S. Mele, hep-ph/9808411;

\bibitem{lattice}
Y. Kuramashi, hep-lat/9910032;
S. Hashimoto, hep-lat/9909136; see also, 
A. Soni, Nucl. Phys. (Proc. Suppl.) {\bf 47}, 43 (1996).
 
\bibitem{Buras}
  G. Buchalla, A. Buras, and M.E. Lautenbacher, Rev. Mod. Phys. {\bf 68},
    1125 (1996).

\bibitem{herrlich}
S. Herrlich and U. Nierste, Nucl. Phys. B {\bf 476}, 27 (1996).

 \bibitem{wolfenstein}
  L. Wolfenstein, Phys. Rev. Lett.  {\bf 51}, 1841 (1983).


\bibitem{sin2beta}
CDF Collaboration, hep-ex/9909003.
%
\bibitem{ciuchini}
	M. Ciuchini, G. Degrassi, P. Gambino and G.F. Giudice,
Nucl. Phys. B {\bf 527}, 21 (1998). 

\bibitem{pdg}
The Particle Data Group, C. Caso {\it et. al.}, Europ. Phys. 
J. {\bf C3}, 1 (1998).

\bibitem{golo}
E. Golowich, in {\em Proceedings of the 2nd International Conference
  on B Physics and CP Violation}, ed. T. Browder {\it et. al.}, 1998, 
hep-ph/9706548; \\
E. Golowich and A. Petrov, Phys. Lett. {\bf B427}, 172 (1998), and
references therein.

\bibitem{nir}
G. Barenboim, G. Eyal, and Y. Nir, hep-ph/9905397; 
G. Eyal and Y. Nir, hep-ph/9908296. 

\bibitem{GL}
M. Gronau and D. London, Phys. Rev. Lett. {\bf 65}, 3381 (1990);
see, however, S. Gardner, Phys. Rev. D {\bf 59}, 077502 (1999). 

\bibitem{cleopenguin}
R. Godang {\it et al.}, (CLEO Collaboration),
Phys. Rev. Lett. {\bf 80}, 3456 (1998).

\bibitem{Kpi}
M. Neubert and J.L. Rosner, Phys. Rev. Lett. {\bf 81}, 5076 (1998).

\bibitem{bcp3}
Searches for direct CP via 
$B^\pm \to \psi/J K^\pm$
are well underway at CLEO, see:
J. Alexander, talk presented at 
{\em the Third International Conference on B physics and CP Violation},
December 3-7, 1999, Taipei, Taiwan. 

\bibitem{a2a1}
M. Neubert and B. Stech, hep-ph/9705292, in 
{\em Heavy Flavours II}, ed. by A.J. Buras and M. Lindner, 
World Scientific, Singapore (1998).

\bibitem{MecaSilva}
C.C. Meca and  J.P. Silva, Phys. Rev. Lett. {\bf 81}, 1377 (1998);
L. Wolfenstein, Phys. Rev. Lett. {\bf 75}, 2460 (1995).

\bibitem{prep}
K. Kiers, A. Soni, and G.-H. Wu, hep-ph/0006280.

\end{thebibliography}
\end{document}